\documentclass[10pt,preprint]{sigplanconf}

\usepackage{amsmath,amssymb}
\usepackage{latexsym}
\usepackage{enumerate}
\usepackage{stfloats}
\usepackage{ctable}
\usepackage[vlined,ruled,linesnumbered]{algorithm2e}
\usepackage{listings}
\usepackage{subfig}
\usepackage{verbatim}
\usepackage{balance}
\usepackage{pslatex}

\usepackage[bookmarks=false,colorlinks, citecolor=black, urlcolor=blue,filecolor=blue, linkcolor=blue]{hyperref}

\newcommand{\cmt}[1]{}

\newcommand{\sectref}[1]{\S\ref{sec:#1}}

\newcommand{\WordToTermMap}{\emph{NLDict}}

\newtheorem{definition}{Definition}

\newtheorem{example}{Example}

\renewcommand{\t}[1]{{\ttfamily #1}}

\newcommand{\ignore}[1]{}

\SetKwInOut{Input}{input}
\SetKwInOut{Output}{output}

\newcommand{\centercell}[1]{\multicolumn{1}{@{\ }c@{\ }|}{#1}}
\newcommand{\head}[1]{\centercell{\bfseries#1}}
\newcommand{\headstart}[1]{\multicolumn{1}{|@{\ }c@{\ }|}{\bfseries#1}}

\newcommand{\domTE}{Text Editing}
\newcommand{\domAut}{Automata}
\newcommand{\domATIS}{ATIS}

\newcommand{\grammar}{G}
\newcommand{\dict}{\ensuremath{D}}
\newcommand{\grule}{R}

\newenvironment{myitemize}{\begin{list}{$\bullet$}
{\setlength{\topsep}{1mm}\setlength{\itemsep}{0.25mm}
\setlength{\parsep}{0.1mm}
\setlength{\itemindent}{0mm}\setlength{\partopsep}{0mm}
\setlength{\labelwidth}{15mm}
\setlength{\leftmargin}{4mm}}}{\end{list}}

\newcommand{\termexp}{t}
\newcommand{\nlinput}{\ensuremath{S}}
\newcommand{\prog}{\ensuremath{P}}
\newcommand{\td}{\mathcal{T}} 
\newcommand{\mapv}{M} 

\newcommand\fVector{\vec{f}}
\newcommand{\featureposone}{f_{\emph{pos1}}}
\newcommand{\featurepostwo}{f_{\emph{pos2}}}
\newcommand{\featurelcaone}{f_{\emph{lca1}}}
\newcommand{\featurelcatwo}{f_{\emph{lca2}}}

\newcommand{\featureorder}{f_{\emph{order}}}
\newcommand{\featureoverlap}{f_{\emph{over}}}
\newcommand{\featuredistance}{f_{\emph{dist}}}

\ifpdf

\else

\fi

\renewcommand{\arraystretch}{1.2}

\newcommand\lang{L}
\newcommand\checker{VC}

\newcommand\terminal{t}
\newcommand\word{\ensuremath{w}}

\newcommand{\bag}{\mbox{\sf\em Bag}}

\newcommand{\covs}{\mbox{\em CoverageScore}}
\newcommand{\maps}{\mbox{\em MappingScore}}
\newcommand{\strs}{\mbox{\em StructureScore}}
\newcommand{\cov}{\mbox{\em\scriptsize cov}}
\newcommand{\map}{\mbox{\em\scriptsize map}}
\newcommand{\str}{\mbox{\em\scriptsize str}}
\newcommand{\classm}{\ensuremath{C_{\map}}}
\newcommand{\classs}{\ensuremath{C_{\str}}}
\newcommand\conn{Conn}
\newcommand\comb{c} 
\newcommand\connProb{\mbox{\sf\em\conn{Probs}}}
\newcommand\weight{\ensuremath{\omega}}

\newcommand{\subprogs}{\mbox{\sf\em SubProgs}}

\newcommand{\UsableWords}{\mbox{\sf\em UsableWords}}
\newcommand{\UsableWordsScr}{\mbox{\scriptsize\sf\em UsableWords}}
\newcommand{\UsedWords}{\mbox{\sf\em UsedWords}}

\newcommand{\lst}{\mbox{\sf\em Likeability}}
\newcommand{\djs}{\mbox{\sf\em Disjointedness}}

\renewcommand{\cite}{\citep*}
\begin{document}

\title{Program Synthesis using Natural Language}

%
%
%
%
%

%
\authorinfo{Aditya Desai}{IIT Kanpur}{adityapd@cse.iitk.ac.in}
\authorinfo{Sumit Gulwani}{MSR Redmond}{sumitg@microsoft.com}
 \authorinfo{Vineet Hingorani\;\;\; Nidhi Jain}{IIT  Kanpur}{viner,nidhij@cse.iitk.ac.in}
\authorinfo{Amey Karkare}{IIT Kanpur}{karkare@cse.iitk.ac.in}
\authorinfo{Mark Marron}{MSR Redmond}{marron@microsoft.com}
\authorinfo{Sailesh \ignore{Kumar Raju} R\;\;\;Subhajit Roy}{IIT Kanpur}{sairaj,subhajit@cse.iitk.ac.in}

\maketitle
\begin{abstract}
Interacting with computers is a ubiquitous activity for millions of people.
Repetitive or specialized tasks often require creation of small, often one-off,
programs. End-users struggle with learning and using the myriad of
domain-specific languages (DSLs)  to effectively accomplish these tasks.

We present a general framework for constructing program synthesizers that take
natural language (NL) inputs and produce expressions in a target DSL. The
framework takes as input a DSL definition and training
data consisting of NL/DSL pairs. From these it constructs a synthesizer by
learning optimal weights and classifiers (using NLP features) that rank the
outputs of a keyword-programming based translation. We applied our
framework to three domains: repetitive text editing, an intelligent tutoring
system, and flight information queries. On 1200+ English descriptions,
the respective synthesizers rank the desired program as the top-1 and top-3 for
80\% and 90\% descriptions respectively.
\end{abstract}

\newcounter{rowno}
\setcounter{rowno}{0}

\begin{table*}[t!]
\scalebox{.87}{
\renewcommand{\arraystretch}{1}
  \begin{tabular}{|@{\ }l@{}|} \hline
    \multicolumn{1}{|c|}{{(a) Grammar}} \\ \hline
    \t{S:= Command | SEQ(Command, Command)} \\
    \t{Command:= ReplaceCmd | RemoveCmd |  InsertCmd | PrintCmd }\\
    \t{ReplaceCmd:= REPLACE(SelectStr, NewString, IterScope)}\\ 
    \t{RemoveCmd:= REMOVE(SelectStr, IterScope)} \\ 
    \t{InsertCmd:= INSERT(PString, Position, IterScope)}\\ 
    \t{PrintCmd:= PRINT(SelectStr, IterScope)} \\
    \t{SelectStr:= (Token, BCond, Occurrence)} \\ 
    \t{IterScope:= (Scope, BCond, Occurrence) | DOCUMENT} \\ 
    \t{Token:= PString | WORDTOK | NUMBERTOK | ...} \\ 
    \t{BCond:= AtomicCond |NOT(AtomicCond)|AND(AtomicCond,AtomicCond)|...} \\ 
    \t{AtomicCond:= STARTSWITH(Token) | CONTAINS(Token) | CommonCond}\\
    \t{CommonCond:= BETWEEN(Token, AnotherToken)| AFTER(Token)| ...}\\ 
    \t{PString:= ConstantString| WORD(ConstantString)|...} \\ 
    \t{Occurrence:= ALL | AtomicOccurrence | ...} \\ 
    \t{AtomicOccurrence:= IntSetByAnd | FirstFew(Integer) | ...}\\
    \t{IntSetByAnd:= Integer | INTSET(IntegerSet)} \\ 
    \t{Scope:= LINESCOPE | WORDSCOPE} \\ 
    \begin{tabular}{@{}ll@{}}
      \t{AnotherToken:= TO(Token)}\rule{17mm}{0pt}& \t{Position:= START | END | ...} \\ 
      \t{NewString:= BY(PString)}                 & \t{ConstantString:= String}\\
      \t{String:= <STRING>}                       & \t{Integer:=<INTEGER>}\\
    \end{tabular} \\ \hline
  \end{tabular}
  \begin{tabular}{|@{}l@{}|} \hline
    \multicolumn{1}{|c|}{{(b) Sample Benchmarks}} \\ \hline
    \begin{tabular}{@{}>{\stepcounter{rowno}\therowno. }c@{}p{7.7cm}l@{}}
      & Remove the first word of lines which start with number. \\
      & Replace ``\&'' with ``\&\&'' unless it is inside ``['' and ``]''. \\ 
      & Add ``\textdollar'' at the beginning of those lines that do not already start with ``\textdollar''. \\  
      & Add ``..???'' at the last of every 2nd statement. \\ 
      & In every line, delete the text after ``//''. \\ 
      & Remove 1st ``\&'' from every line. \\ 
      & Add the suffix ``\_IDM'' to the word right after ``idiom:''. \\ 
      & Delete all but the 1st occurrence of ``Cook''. \\ 
      & Delete the word ``the'' wherever it comes after ``all''. \\ 
      & Print data between ``$<$url$>$'' and ``$<$/url$>$''. \\ 
    \end{tabular}
    \\ \hline
    \multicolumn{1}{c}{} \\\hline
    \multicolumn{1}{|c|}{{$\!\!$(c) $\!\!$Variations in NL for description of the same task.}}
    \setcounter{rowno}{0}\\ \hline
    \begin{tabular}{@{\ }>{\stepcounter{rowno}\therowno.\ \  }c@{}p{7.7cm}}
      & Prepend the line containing ``P.O. BOX'' with ``*'' \\
      & Add a ``*'' at the beginning of the line in which the string "P.O. BOX" occurs\\
      & Put a ``*'' before each line that has ``P.O. BOX''  in it\\
      & Put ``*'' in front of lines with ``P.O. Box'' as a substring\\
      & Insert ``*'' at the start of every line which has ``P.O. BOX'' word
    \end{tabular}        \\ \hline
  \end{tabular}
}
\caption{Grammar and sample benchmarks for the Text Editing domain.}
\label{tab:textediting}
\end{table*}

\section{Introduction}
{\em Program synthesis} is the task of automatically synthesizing a program
in some underlying {\em domain-specific language (DSL)} from
a given specification~\cite{dimensions}. 
Traditional program synthesis, that of synthesizing 
programs from {\em complete
  specifications}~\cite{deductive80,
  popl10,loopfree:pldi11,armando:asplos06}, have not yet seen
wide adoption as it is often difficult to automatically 
check that a specification is satisfied by the synthesized program.
More significantly, these specifications are difficult to write. 

Recent work has experimented with another class of ({\em
  possibly incomplete}) specifications, namely {\em
  examples}~\cite{lieberman,cypher-pbd,synasc12,popl11,cacm12}.
Programming by Example (PBE) systems have seen much wider
adoption, thanks to the ease of providing such a
specification.  However, they are not ideal for specifying
certain kinds of operations such as {\em filter} or {\em
  reduce}. In particular, conditional operations generally
require examples exercising both the true and false paths
leading to rapid growth in the number of examples needed.
The classic L* algorithm~\cite{Angluin87}, a PBE system
for describing a regular language, has the well-known
drawback of requiring too many examples.  Even
state-of-the-art PBE systems like FlashFill~\cite{flashfill}
are limited in their handling of conditionals. {Moreover, in
  such tasks, PBE requires analyzing the contiguous chunk of
  input data on which edits have been performed. This leads
  to scalability issues on text files, which are much larger
  than strings.  (FlashFill is restricted to work with
  strings with at most 256 characters). } Describing tasks
using examples have other drawbacks: For some domains like
air travel information systems (ATIS), it is not even clear
what an ``example'' is.  It turns out that operations like
filter and reduce, and their compositions can be specified
much more easily and concisely using natural language~(NL).

In this paper, we address the problem of synthesizing
programs in an underlying DSL from NL. NL 
is inherently imprecise; hence, it may not be
possible to guarantee the correctness of the synthesized
program. Instead, we aim to generate a ranked set of programs
and let the user possibly select one of those programs by
either inspecting the source code of the program, or the
result of executing that program on some test inputs.
This user interaction is similar
to what is employed in any PBE technology like Flash
Fill. The reader may also liken this process to how users
search for and select their desired results in a search
engine.  Further, similar to a search engine, the synthesis
algorithm in this paper is able to consistently produce and
rank the desired result program in the top spot, over 80\% of
the time, or in the top 3 spots, over 90\% of the time in
our benchmarks. To give users confidence in the program they 
choose, we show both the translation of the code into 
disambiguated English and/or run it to show the result as a preview. 
 
Unlike most of the existing synthesis techniques that specialize to a specific 
DSL, our methodology can be 
applied to a variety of DSLs. Our methodology
requires two inputs from the synthesis designer: 
(i) The DSL definition,
(ii) Training data consisting of example pairs of English sentences and 
corresponding intended programs in the DSL. A training phase infers 
a {\em dictionary} relation over pairs of English words and DSL terminals 
(in a semi-automated interactive manner), and 
optimal weights/classifiers (in a completely automated manner) 
for use by the generic synthesis algorithm. Our 
approach can be seen as a meta-synthesis framework for constructing 
NL-to-DSL synthesizers.

The generic synthesis algorithm (Alg.~\ref{algo:synth}) takes
as input an English sentence and generates a ranked set of
likely programs. First, it uses a {\em bag algorithm}
(Alg.~\ref{algo-algoa}) to efficiently compute the set of all
{\em consistent} DSL programs whose terminals are related to
the words that occur in the sentence. For this, it uses a
{\em dictionary} (learned during the training phase) that is
a relation over English words and DSL
terminals. Then, it ranks these programs based on a set of
scoring functions (\sectref{ranking}). These functions are
inspired by our view of the Abstract Syntax Tree (AST) of a
program as involving two constituents: the set of terminals
in the program, and the tree structure between those
terminals.  We use a weighted linear combination of 3 scores
to determine the rank of each program: (i) a {\em coverage
  score} that captures the intuition that results that ignore
many words in the user input are unlikely to be correct (ii)
a {\em mapping score} that captures the intuition that
English words can have multiple meanings and intended actions
wrt. the DSL but we prefer the more probable
interpretations (iii) a {\em structure score} that uses the
insight that both natural language and programming languages
have common idiomatic
structures~\cite{naturalnessofsoftware}, and prefer
the more \emph{natural} results.

The classifiers to compute these scores, as well as the
weights for combining the scores are learned during the
training phase using off-the-shelf machine learning
algorithms.  The novelty of of our approach lies in the
generation of training data for classifier learning from the
top-level training data (Alg.~\ref{algo:comp2Train}
and~\ref{algo:comp3Train}), and in smoothing a discrete
scoring metric into a continuous and differentiable loss
function for effective learning of weights
(\sectref{learning-combination}).

This paper makes the following contributions:
\begin{myitemize}
\item We describe a meta-synthesis framework for
  constructing NL-to-DSL synthesizers, consisting of a
  synthesis algorithm (\sectref{algorithms}) for
  translating English sentences into corresponding programs
  in the underlying DSL, and a training phase for learning a
  dictionary and weights that are used by the
  synthesis algorithm (\sectref{learning}).  Our methodology
  can be applied to new DSLs, and requires only the 
  DSL definition along with translation pair training data.

\item We apply our generic framework to three different domains, 
  namely automating end-user data manipulation (\sectref{text}),
  generating problem descriptions in intelligent tutoring systems
  (\sectref{automata}), and database querying (\sectref{atis}). 
  In cases where comparisons can be made with state-of-the-art 
  NLP based approaches, the results of the approach presented 
  in this paper are competitive. 

\item We gather an extensive corpus of data consisting of
  1272 pairs of English descriptions and corresponding
  programs.  We use this data for evaluation in this paper,
  and provide it as a resource for researchers in the
  community.  Of these, 535 English descriptions come from
  the {\em Air Travel Information System (ATIS)} benchmark
  suite, 227 come from another corpus~\cite{acl11}, while 510
  English descriptions were collected by us from various
  online sources (including help forums and course
  materials), textbooks, and user studies.

\item We evaluate the effectiveness of our approach on 3
  different DSLs (\sectref{results}). The NL-to-DSL synthesizers
  produced by our framework run in $1-2$ seconds on average per
  benchmark and produce a ranked set of candidate programs with the
  correct result in the \mbox{top-1}/top-3 choices for over 80\%/90\%
  benchmarks respectively.
\end{myitemize}

\section{Motivating Scenarios}\label{sec:motivation}
We describe 3 different domains where a
NL-to-DSL synthesizer is useful: text
editing (\sectref{text}), automata construction problems for
intelligent tutoring (\sectref{automata}), and answering
queries for an air travel information systems
(\sectref{atis}).  
\subsection{Text Editing (End-User Programming)}
\label{sec:text}



Through a study of help forums for Office suite applications like
Microsoft Excel and Word, we observed that users frequently request
help with repetitive {\em text editing} operations such as insertion,
deletion, replacement, or extraction in text files.  These operations
(\autoref{tab:textediting}(b)) 
are more complicated than simple search-and-replace of a
constant string by another in two ways.  First, the string being
searched for is often not constant and instead requires regular
expression matching. Second, the editing is often conditional on the
surrounding context. Programming of even such relatively simple tasks
requires the user to understand syntax
and semantics of regular expressions, conditionals, and loops, which
are beyond the ability of most end-users.

This inspired us to design a command language for
text-editing (a subset of the grammar is shown in
\autoref{tab:textediting}(a)) that includes key commands
\t{Insert}, \t{Remove}, \t{Print} and \t{Replace}. Each of
these commands relies on an \t{IterScope} expression that
specifies the region (a set of lines, a set of words, or the
entire Word document) that the text editing operation is on.
The \t{SelectStr} production includes a \t{Token}, which
allows for limited wild-card matching (e.g., an entire
\t{WORDTOK}, \t{NUMBERTOK}, or a pattern specified by
\t{PString}), a Boolean condition \t{BCond} that acts as an
additional (local) filter on the matched value, and an
\t{Occurrence} value that performs an index based selection
from the resultant matches. Use of the occurrence values like
\t{FirstFew}(N) (from \t{AtomicOccurrence}) when performing a
\t{Remove} results in the removal of \emph{only} the first N
items (here N is a positive integer) that match the
condition, while use of \t{ALL} will instead result in all
matches of the condition being removed.  The Boolean
conditions \t{BCond} cover the standard range of string
matching predicates (\t{CONTAINS}, \t{STARTSWITH} etc.) and
allow conjunction of conditions (\t{AND}, \t{NOT} etc.).  The
\t{CommonCond} production specifies the position relative to
the string token(s) that occurs after it (\t{AFTER}), before
it (\t{BEFORE}), or around it (\t{BETWEEN}) acts as another
(global) filter.
{\autoref{tab:textediting}(c) describes a sample of the variations that our system can handle for description of a task that is expressible in our DSL.
}

\begin{example}
For the text editing task described in task 1 in
\autoref{tab:textediting}(b), our system produces the
following translation:\\[1mm]
\small
\noindent\t{REMOVE(SelectStr(WORDTOK, ALWAYS, INTEGER(1)),
  IterScope(LINESCOPE, STARTSWITH(NUMBERTOK), ALL))}
\end{example}

\begin{example}
Given the English description for task 2 in
\autoref{tab:textediting}(b) our system produces the
following translation: \\[1mm]
\small
\noindent\t{REPLACE(SelectStr(STRING($\&$),
  NOT(BETWEEN(\linebreak STRING([), TO(STRING(])))), ALL),
  BY(STRING(\linebreak $\&\&$)), DOCUMENT))}
\end{example}

Our belief is that once users are able to accomplish these
types of smaller conditional and repetitive tasks, they can
easily accomplish other more complex tasks by reducing them
to a sequence of smaller tasks.

\subsection{Automata Theory (Intelligent Tutoring)}
\label{sec:automata}
\begin{table}[t]
\setcounter{rowno}{0}
\scalebox{.93}{
\renewcommand{\arraystretch}{1}
\begin{tabular}{|@{}>{\ \stepcounter{rowno}\therowno.\ \  }c@{}p{8.3cm}|}
\hline
& Consider the set of all binary strings where the difference between the number of ``0'' and the number of ``1'' is even.\\
& The set of strings of ``0'' and ``1'' such that at least one of the last 10 positions is a ``1''.\\
& the set of strings w such that the symbol at every odd position in w is ``a''.\\
& Let L1 be the set of words w that contain an even number of ``a'', let L2 be the set of words w that end with ``b'', let L3 = L1 intersect L2. \\
& The set of strings over alphabet 0 to 9 such that the final digit has not appeared before.\\
\hline
\end{tabular}}
\caption{Sample benchmarks for the Automata  domain.}
\label{tab:automata}
\end{table}
 Results from formal methods research have been used
in many parts of intelligent tutoring
systems~\cite{education13} including problem generation,
solution generation, and especially feedback generation for a
variety of subject domains including
geometry~\cite{geometry:pldi11} and automata
theory~\cite{ijcai13:grading}.  Each of these domains
involves a specialized DSL that is used by a problem generator tool
to create new problems, a solution generation tool to produce
solutions, and more significantly, a feedback
generation tool to provide feedback on student solutions.

Consider the domain of automata constructions, where students
are asked to construct an automaton that accepts a 
language whose description is provided in English (For some examples, see
\autoref{tab:automata}).  We designed
a DSL based on the description provided by Alur
et.al.~\cite{ijcai13:grading} on constructs required to
formally specify such languages. This DSL contains predicates
over strings, Boolean connectives, functions that return
positions of substrings, and universal/existential
quantification over string positions.  As stated
in~\cite{ijcai13:grading}, such a language is used to
generate feedback for students' incorrect attempts in two
ways: (i) it is used by a solution generation tool to
generate correct solutions against which a student's
attempts are graded, (ii) it is also used to
provide feedback and generate problem variations consistent with a student's
attempt. This feedback generation tool has been deployed
in the classroom and has been able to assign grades and
generate feedback in a meaningful way while being both faster
and more consistent than a human. 
Our synthesis methodology can be
used to automatically generate the specifications needed by this system from
natural language descriptions.

\begin{example}
Specification 1 in \autoref{tab:automata} is translated as: \\
\small
\t{ISEVEN(DIFF(COUNT(STRING(0)), COUNT(STRING(1)\\
   \ )))}
\end{example}

\begin{example}
Specification 2 in \autoref{tab:automata} is translated as:\\
\small
\t{EXISTSINT(LASTFEW(INTEGER(10)), STREQUALS(\\
  \ SYMBOLATP(), STRING(1)))}
\end{example}

\subsection{Air Travel Information Systems (ATIS)} 
\label{sec:atis}%
\begin{table}[t]
\setcounter{rowno}{0}
\scalebox{.93}{
\renewcommand{\arraystretch}{1}
\begin{tabular}{|@{}>{\ \stepcounter{rowno}\therowno.\ \  }c@{}p{8.3cm}|}
\hline
& I would like the time of your earliest flight in the morning from Philadelphia to Washington on American Airlines. \\
& I need information on a flight from San Francisco to Atlanta that would stop in Fort Worth. \\ 
& What is the earliest flight from Washington to Atlanta leaving on Wednesday September fourth. \\ 
& Okay we're going from Washington to Denver first class ticket I would like to know the cost of a first class ticket. \\
& What ground transportation is there from the airport in Atlanta to downtown. \\ 
\ignore{
& i would like to know if i fly on american flight number 813 from boston to oakland if i will stop enroute at another city \\ 
& show me the flight from detroit to westchester county with the highest one way fare \\  
& i would like to make a round trip between washington and san francisco \\ 
& i would like to know what type of aircraft will be used on the morning of july seventh from atlanta to boston \\ 
& i want to know the time of the latest flight i can take from washington to san francisco where i can get a dinner meal \\
}
\hline
\end{tabular}
}
\caption{Sample benchmarks for the ATIS domain.}
\label{tab:atis}
\end{table}

ATIS is a standard benchmark for querying air travel
information, consisting of English queries and an associated
database containing flight information. It has long been used
as a standard benchmark in both natural language processing
and speech processing communities. \autoref{tab:atis} shows
some sample queries from the ATIS suite. For ATIS, we designed a DSL that is based
around SQL style row/column operations and provided support for 
predicates/expressions that correspond to important concepts
in air-travel queries, arrival/departure locations,
times, dates, prices, etc. 

\begin{example}
The first query in the ATIS examples, \autoref{tab:atis}, is translated into our DSL as:\\
\small
  \t{ColSet(AtomicColSet(DEP\_TIME()), ROW\_MIN(\\
  \ DEP\_TIME(), AtomicRowPredSet(AtomicRowPred(\\
  \ EQ\_DEP(CITY(philadelphia)), Unit\_Time\_Set(\\
  \ TIME(morning)), EQ\_ARR(CITY(washington)),\\
  \ EQ\_AIRLINES(AIRLINES(american))))))}
\end{example}

\section{Problem Definition}
\label{sec:problemDef}

We study the problem of synthesizing NL-to-DSL synthesizers,
given a DSL definition and example training data. A DSL
$\lang = (\grammar, \checker)$ consists of a context free
grammar $\grammar$ (with terminal symbols denoted by
$\grammar_T$ and production rules denoted by $\grammar_R$),
and a syntactic/semantic checker $\checker$ that can check
whether or not a given program belongs to the grammar
$\grammar$ and is semantically meaningful.  The {\em training
  data} consists of a set of pairs $(\nlinput,\prog)$, where
$\nlinput$ is an English sentence and $\prog$ is the
corresponding intended program from the DSL $\lang$. A
sentence is simply a sequence of words $[\word_1, \word_2,
  \ldots, \word_n]$.  The goal of the generated NL-to-DSL
synthesizer is to translate an English sentence to a ranked
set of programs, $[\prog_1, \prog_2, \ldots, \prog_k]$, in
$\lang$.

\newcommand{\gformula}{P}
\newcommand{\hole}{\Box} 
\newcommand{\used}{\emph{UsedW}}
\newcommand{\combination}{\emph{SubAll}}
\newcommand{\myquad}{\ \ }

\section{NL to DSL Synthesis Algorithm}
\label{sec:algorithms}

Our synthesis algorithm (\autoref{algo:synth}) takes a natural language command from the user 
and creates a ranked list of candidate DSL programs. The first step (loop on \autoref{line:initloop}) is to convert each of the words in the user
input into one or more terminals (function names or values) using the NL to program 
terminal Dictionary $\WordToTermMap$. This loop ranges over the length of the input sentence and 
for each index looks up the set of terminals in the DSL that are associated with the word at that index in
$\WordToTermMap$. Fundamentally, $\WordToTermMap$ encodes, 
for each terminal, which English language words are likely to indicate the presence of that 
terminal in the desired result program. This map can be constructed in a semi-automated 
manner (\sectref{dictcons}). Once this association has been made for a terminal $t$ 
we store a tuple of the terminal and a singleton map, relating the index of the word to a 
terminal that was produced, into the set $R_0$
(\autoref{line:inittuple}).

For each natural language word, the dictionary
$\WordToTermMap$ associates a
set of terminals with it. The terminals may be constant
values or function applications with {\em
  holes} ($\hole$) as arguments.  Thus, algorithm
(\autoref{algo:synth}) (when applying NLDict on \autoref{line:dictapply}) can
create incomplete programs, where some arguments to functions
are missing. For example, Consider the sentence ``Print all lines that do not
contain 834''. Since the grammar contains {\tt PrintCmd :=
PRINT(SelectStr, IterScope)} as a production and the
dictionary relates the word ``print'' to the function {\tt PRINT},
the partial program {\tt PRINT($\hole$, $\hole$)} will be
generated. These holes are later replaced by other programs
that match the argument types SelectStr and IterScope. 

Once the base set of terminals has been constructed, the
algorithm uses the $\bag$ algorithm (\autoref{algo-algoa}) to
generate the set of all {\em consistent} programs,
$\emph{Res}_T$, that can be constructed from it (\autoref{line:bag}). The final
step is to rank (\sectref{ranking}) this set of programs,
using a combination of scores and weights, in the loop on
\autoref{line:scst}.
\begin{algorithm}[t]
\small
\caption{NL to DSL Synthesis Algorithm \label{algo:synth}}
\KwIn{NL sentence $\nlinput$, Word-to-Terminal dictionary $\WordToTermMap$}
\KwOut{Ranked set of programs}
 $R_0 \gets \emptyset$\;
 \For{$i \in \lbrack 0, S.\text{Length} - 1 \rbrack$ \label{line:initloop}}{
    $T \gets \WordToTermMap(S\lbrack i \rbrack$)\; \label{line:dictapply}
    \lForEach{$t \in T$}{
      $R_0 \gets R_0 \cup (t, \emph{SingletonMap}(i, t))$ \label{line:inittuple}
    }
 }
 $\emph{Res}_T \gets \bag(\nlinput, R_0)$\; \label{line:bag}
$\emph{Res}_{\prog} \gets \{ \prog \vert \exists \mapv \text{ s.t. } (\prog, \mapv) \in \emph{Res}_T \}$\;
\lForEach{program $\prog \in \text{Res}_{\prog}$}{$\emph{score}(\prog) \gets -\infty$}
 \ForEach{program $(\prog, \mapv) \in \text{Res}_T$}
{\label{line:scst} 
      $s_{\cov} \gets \covs(\prog,\nlinput,\mapv) \times \weight_{\cov}$\;
      $s_{\map} \gets \maps(\prog,\nlinput,\mapv) \times \weight_{\map}$\;
      $s_{\str} \gets \strs(\prog,\nlinput,\mapv) \times \weight_{\str}$\;
      $\emph{score}(\prog) \gets \max(\emph{score}(\prog),  s_{\cov} + s_{\map} + s_{\str}$)\;

\label{line:scen}
}
\Return{set of programs in $\text{Res}_{\prog}$ ordered by $score$}
\end{algorithm}
\subsection{Synthesizing Consistent Programs}
\label{sec:bag}

A program $\gformula$ in the DSL is either an atomic value (i.e., a terminal in \grammar), 
or a function/operator applied to a list of arguments. By convention we represent function application 
as \emph{s-expressions} where a function $F$ applied to $k$ arguments is written 
$(F, P_1, \ldots, P_k)$.

\paragraph
{Consistent Programs and Witness Maps.}
Given a DSL $\lang = (\grammar, \checker)$, we say a program
$\prog$ in language $\lang$ is consistent with 
a sentence $\nlinput$ if there exists a map $\mapv$ that maps (some) word occurrences in $\nlinput$ to terminals in
$\grammar_T$ such that the range of $\mapv$ equals the set of terminals in
the program $\prog$.\footnote{Since the same English word can occur at
  different positions in $\nlinput$, having different
  meanings, any map $\mapv$ must take the position information
  also as an argument. For simplicity of exposition, we ignore this in the
  paper.} We refer to such a map $\mapv$ as a {\em witness} map, 
and use the notation \t{WitnessMaps}$(\prog,\nlinput)$ to denote 
the set of all such maps.

\paragraph{Usable and Used Words.}
Let $\nlinput$ be an English sentence, $\prog$ be a program consistent with $\nlinput$, 
and $\mapv$ be any witness map.
$\UsableWords(\nlinput)$ are those word occurrences in $\nlinput$ that are mapped
to some grammar terminal and hence might be useful in translation. 
$\UsedWords(\nlinput,\mapv)$ is the set of 
usable word occurrences in $\nlinput$ that are used as part of the map
$\mapv$. 
\begin{eqnarray*}
 \UsableWords(\nlinput) & = & \{i \mid S \lbrack i \rbrack \in \emph{Domain}(\WordToTermMap) \}\\
 \UsedWords(\nlinput, \mapv) & = &  \UsableWords(\nlinput) \cap \emph{Domain}(\mapv)
\end{eqnarray*}

\paragraph{Partial Programs.}
A partial program extends the notion of a program to also
allow for a {\em hole} ($\hole$) as an argument. A hole is a
symbolic placeholder where another complete program (program
without any hole) can be placed to form a larger program.  To
avoid verbosity, we often refer to a {\em partial program} as
simply a {\em program}.

Given a partial program $\gformula = (F, \ldots, \hole, \ldots)$
with a hole $\hole$, we can substitute a complete program $\gformula'$ to 
fill the hole: 
\[
\gformula\lbrack \hole \gets \gformula'\rbrack = \left\{
\begin{array}{cl}
(F, \ldots, \gformula', \ldots) & \text{if } \checker((F, \ldots, \gformula', \ldots)) \\
\bot & \text{otherwise}\\
\end{array}
\right.
\]
The validity check, $\checker$, ensures that all synthesized programs 
are well defined in terms of the DSL grammar and type system (otherwise 
we return the {\em invalid} program $\bot$).

\begin{algorithm}[t!]
\small
\caption{Bag} 
\label{algo-algoa}
\KwIn{NL sentence $\nlinput$, Initial Tuple Set $B_0$} 
$\emph{result} \gets B_0$\; \label{L-algoa-smaller}
\Repeat{\mbox{$\text{oldResult} = \text{result}$}}{\label{L-algoa-repeat}
  $\emph{oldResult} \gets \emph{result}$\; 
  \ForEach{$(\gformula_1, \mapv_1), (\gformula_2, \mapv_2) \in \text{ result}$}{
    $\emph{okpc} \gets \gformula_1 \text{ is partial } \land \gformula_2 \text{ is complete }$\;\label{L-algoa-complete}
    $\emph{disjoint}\!\!\gets\!\!\UsedWords(\nlinput,\mapv_1)~\cap~\UsedWords(\nlinput,\mapv_2)=\emptyset$\;\label{L-algoa-intersect}
    \If{$\text{okpc} \land \text{disjoint}$}{  
      $\emph{combs} \gets \combination(\gformula_1,\gformula_2) - \{ \bot \}$\; \label{L-algoa-rep}
      $\emph{new} \gets \{(\gformula_r, \mapv_1 \cup \mapv_2) \vert \gformula_r \in \emph{combs} \}$\; \label{L-algoa-add}
    }
  }
  $\emph{result} \gets \emph{result} \cup \emph{new}$\;
}
return $\emph{result}$\;
\end{algorithm}

\paragraph{Combination.}
The combination operator $\combination$ generates the set of all
programs that can be obtained by substituting a complete program
$\gformula'$ in some hole of a partial program $\gformula$. 
This is done by going over all the arguments of $\gformula$ and producing substitutions for 
argument positions with holes.
Given partial program $\gformula = (F, \gformula_1, \ldots, \gformula_k)$ and complete program 
$\gformula'$, we have:
\[
\combination(\gformula, \gformula') = \{ \gformula\lbrack \hole_i \gets \gformula' \rbrack \vert \gformula_i = \hole_i \} 
\]

\paragraph{Bag Algorithm.}
The $\bag$ algorithm (\autoref{algo-algoa}) is based on computing the closure of a set of programs by 
enumerating all possible well-typed combinations of the programs in the set. The main loop (line~\ref{L-algoa-repeat}) is a fixpoint 
iteration on the \emph{result} set of programs that have been constructed. 

The requirement that $P_2$ is a complete program (line~\ref{L-algoa-complete}) when applying the $\combination$ function ensures that 
the only holes in the result programs are holes that were originally in $P_1$. We restrict the initialization of 
$B_0$ to include only complete programs and partial programs with holes at the top level only. Using this restriction 
we can inductively show that at each 
step all partial programs only have holes at the top-level. Thus, we can efficiently compute the fixpoint of all
possible programs in a bottom-up manner.
The condition $\UsedWords(\nlinput, \mapv_1) \cap
\UsedWords(\nlinput, \mapv_2) = \emptyset$ (line~\ref{L-algoa-intersect}) ensures that the two programs 
do not use overlapping sets of words from the user input. This ensures that the final program cannot create 
multiple sub-programs with different meanings from the same part of the user input. This also ensures that
the set of possible combinations has a finite bound based on the number of words in the input. 
Line~\ref{L-algoa-rep} constructs the set of all possible substitutions of $\prog_2$ into holes in $\prog_1$ (ignoring 
any invalid results). For each of the possible substitutions we add the result (and the union of the $\mapv$ 
maps) to the $\emph{new}$ program set (\autoref{L-algoa-add}). Since \cmt{we previously checked that} the domains of the maps 
were disjoint the union operation is well-defined.

The $\bag$ algorithm has a high
recall, but, in practice, it may generate spurious programs
that arise as a result of arbitrary rearrangement of the
words in the English sentence. To account for this, the
correct translation is reported by selecting the top-most
rank program based on features of the program and the parse tree 
of the sentence.

\subsection{Ranking Consistent Programs}
\label{sec:ranking}

We view the abstract syntax tree of the synthesized program as
consisting of two important constituents: the set of terminals in the
program, and the tree structure between those terminals. We use these
constituents to compute the following three scores to determine the
rank of a consistent program: (i) a {\em coverage score} that reflects
how many words in the English sentence were mapped to some operation or value 
in the program, (ii) a {\em mapping score} that reflects the likelihood that a 
word-to-terminal mapping is capturing the user intent (iii) a 
{\em structure score} that captures the naturalness of the tree program structure and 
the connections between parts of the program and the parts of the sentence 
that generated them. 


\subsubsection{Coverage Score}
For a given sentence $\nlinput$, a candidate translation $\prog$, and a witness map $\mapv$, the 
coverage score is defined as,
\[
  \covs(\prog, \nlinput, \mapv) = 
  \begin{array}{c}
    \left| \UsedWords(\nlinput, \mapv) \right| \\ \hline 
    \left| \UsableWords(\nlinput) \right|
  \end{array}
\]
The $\covs(\prog, \nlinput, \mapv)$ denotes the
fraction of available information in $\nlinput$ that is actually used to generate
$\prog$. Intuitively we want to prefer programs that make more use 
of the information provided by the user input

\begin{example}
Consider  possible translations for an input $\nlinput$:

\noindent{\small
$\begin{array}{@{}r@{}cl}
\nlinput &:& \mbox{\t{find the cheapest flight from Washington}}\\
         & & \mbox{\t{to Atlanta}}\\
\prog_1  &:& \mbox{\t{MIN\_F(COL\_FARE(), AtomicRowPredSet(EQ\_DEP(}}\\
         & & \quad \mbox{\t{CITY(Washington)), EQ\_ARR(CITY(Atlanta))))}}\\
\prog_2  &:& \mbox{\t{AtomicRowPredSet(EQ\_DEP(CITY(Washington)),}} \\
         & & \quad\mbox{\t{EQ\_ARR(CITY(Atlanta)))}}
\end{array}
$}\\
The first program $\prog_1$ makes use of all parts of the user input, including the desired 
cheapest fare, while the second program $\prog_2$ ignores this information. The Coverage score 
enables us to rank $\prog_1$ higher than $\prog_2$. 
\end{example}

\subsubsection{Mapping Score}
For any word $\word$ there may be multiple terminals
(functions or values) in the set $\WordToTermMap(\word)$ each
of which corresponds to a different interpretation of
$\word$. We use machine learning techniques to obtain a
classifier $\classm$ based on the part-of-speech (POS) tag
provided for the word by the Stanford NLP
engine~\cite{stanfordnlp}.  $\classm.\text{Predict}$ function
of the classifier predicts the probability of each
word-to-terminal mapping being correct.  We use predictions
from \classm\ to compute the \maps, the likelihood that
terminals in $\prog$ are correct interpretation of
corresponding words in $\nlinput$.\\[1mm]
\noindent$\maps(\prog, \nlinput, \mapv) =$\\[-4mm]
\[\qquad\qquad\prod_{\word\in \UsableWordsScr(\nlinput)} 
  \classm.\text{Predict}(\word, \text{POS}(\word, \nlinput), \mapv(\word))\]

A limitation of the \maps\ score is that it looks only at the
mapping of a word but not its relation to other words and how
they are are mapped by the translation.  Thus, interchanging
a pair of terminals in a correct translation gives us an
incorrect translation which has the same score.

{
\begin{example}
Consider the input $\nlinput$ and two possible translations: \\

\noindent{\small
$\begin{array}{@{}r@{}cl}
\nlinput &:& \mbox{\t{If ``XYZ'' is at the beginning of the line,}} \\
         & &  \mbox{\t{replace ``XYZ'' with ``ABC''}} \\
\prog_1  &:& \mbox{\t{REPLACE(SelectStr(STRING(XYZ), ALWAYS(),}}\\
         & & \quad \mbox{\t{ALL()), BY(STRING(ABC)), IterScope(}}\\
         & & \quad \mbox{\t{LINESCOPE(), STARTSWITH(STRING(XYZ)),}}\\
         & & \quad \mbox{\t{ALL()))}}\\
\prog_2  &:& \mbox{\t{REPLACE(SelectStr(STRING(XYZ), ALWAYS(),}} \\
         & & \quad \mbox{\t{ALL()), BY(STRING(ABC)), IterScope(}}\\
         & & \quad \mbox{\t{LINESCOPE(), BEFORE(STRING(XYZ)), ALL()))}}
\end{array}
$}\\

Both the programs use same sets words, so they have the same
coverage score.  The only difference is that the word
``beginning'' is mapped to STARTSWITH (POS: Verb Phrase) in
$\prog_1$, and to BEFORE (POS: Prepositional Phrase) in
$\prog_2$. Mapping score helps in identifying $\prog_1$ as
the correct choice.

\end{example}
}

\subsubsection{Structure Score}
Structure score captures the notion of naturalness in the placement of
sub-programs. We use connection features obtained from the sentence 
$\nlinput$, the natural language parse tree for the sentence,
NLParse($\nlinput$), and the corresponding program $\gformula$ 
to define the overall structure score. These features are used to 
produce the classifier $\classs$ which computes the probability that 
each of the combinations in $\prog$ is correct. 

\begin{definition}[Connection]
For a production $\grule \in
\grammar_R$ of the form $ N \rightarrow N_1\ldots N_i\ldots N_j\ldots
N_k$, the tuple $(\grule, i, j)$ where $1\leq i, j \leq
k,$ and $i\neq j$ is called a connection.
\end{definition}

\begin{definition}[Combination]
Consider the program $\prog$ $=$ $(\prog_1,
\prog_2,$ $\ldots,$ $\prog_k)$ generated using the production $\grule: N\rightarrow
N_1N_2\ldots N_k$, such that $N_i$ generates $\prog_i$ for $1
\leq i \leq k$. We say the pair of sub-programs $(\prog_i, \prog_j)$ is
combined via connection 
$(\grule, i, j)$
and this combination is
denoted as $\conn(\prog_i, \prog_j)$.
\end{definition}

The overall \strs\ is obtained by taking the geometric mean
of the various connection probabilities of the scores for the
program $P$---this normalizes the score to account for
programs with differing numbers of connections.
\[
\begin{aligned}
&  \strs(\prog, \nlinput, \mapv) \!=\!  \mbox{GeometricMean}(\connProb(\prog, \nlinput, \mapv)) \\
%
&  \connProb(\prog, \nlinput, \mapv)\! = \!\!\!\!\!\!\!\! 
  \bigcup_{\conn(\prog_i, \prog_j) \mbox{ in } \prog}
  \{\classs[\conn].\text{Predict}(\fVector, 1)\} \\
& \ \ \mbox{ where } \fVector \equiv \langle       \featureposone, \featurepostwo, \featurelcaone, \featurelcatwo, \featureorder, \featureoverlap, \featuredistance \rangle\\
& \ \ \qquad\qquad\quad\mbox { computed for $\prog_i$ and $\prog_j$ using $\prog$, \nlinput, and \mapv.}
\end{aligned}
\]

We obtain separate classifier, \classs[\conn], for each connection \conn. The function \classs[\conn].\text{Predict} asks the classifier to predict the probability that f-vec belongs to class 1 (i.e., present in correct translation). The other class is 0.

\newcommand\spanw{\psi}

Given a program $\prog$ and input sentence $\nlinput$ that are related by a witness map $\mapv$ and 
the parse tree NLParse($\nlinput$), the following functions define several useful relationships:
\[
\begin{aligned}
&\text{TreeCover}(\prog, \nlinput, \mapv) = \text{minimal sub-tree } T_{\text{sub}} \text{ of } \text{NLParse}(\nlinput) \\
&\ \ \text{s.t. } \UsedWords(\nlinput, \mapv) \subseteq \UsableWords(T_{\text{sub}})\\
&\text{Root}(\prog, \nlinput, \mapv) = \text{root node of } \text{TreeCover}(\prog, \nlinput, \mapv)\\
&\text{Span}(\prog, \mapv) = \left[\text{Min}(\emph{Domain}(\mapv)), \text{Max}(\emph{Domain}(\mapv))\right]
\end{aligned}
\]

In the rest of the section, we assume that $\prog_1$ and $\prog_2$ denote two
sub-programs of $\prog$. The following features determine the naturalness of the connections between 
$\prog$, $\prog_1$, $\prog_2$, and $S$ :

\begin{definition}[Root POS Tags]
Part-of-speech features are the POS tags
assigned by the NL Parser to the root nodes of the sub-trees associated
with $\prog_1$ and $\prog_2$ respectively:
\[
\begin{aligned}
&\featureposone \equiv \text{POS}(\text{Root}(\prog_1, \nlinput, \mapv))\\
&\featurepostwo \equiv \text{POS}(\text{Root}(\prog_2, \nlinput, \mapv))\\
\end{aligned}
\]
\end{definition}

The features $\featureposone$ and $\featurepostwo$ help to learn the
phrases that are commonly combined using a particular connection.

\begin{definition}[LCA Distances]
Let LCA be the least-common-ancestor of $\text{Root}(\prog_1,
\nlinput, \mapv)$ and $\text{Root}(\prog_2, \nlinput, \mapv)$.  The
least-common-ancestor distance features are the tree-distances from
LCA to the root nodes of the sub-trees associated with $\prog_1$ and
$\prog_2$ respectively:
\[
\begin{aligned}
&\featurelcaone \equiv \text{TreeDistance}(\text{LCA},
  \text{Root}(\prog_1, \nlinput, \mapv))\\ 
&\featurelcatwo \equiv
  \text{TreeDistance}(\text{LCA}, \text{Root}(\prog_2, \nlinput, \mapv))
\end{aligned}
\]
\end{definition}

\begin{definition}[Order]
The order feature is determined by the positions of the roots 
of the sub-tree roots associated with $\prog_1$ and $\prog_2$ in the 
\emph{in-order} traversal of NLParse(\nlinput).
\[
\featureorder = \left\{
\begin{array}{rl}
1 & \ \ \mbox{if }\text{Root}(\prog_1, \nlinput, \mapv) \text{ occurs  before } \text{Root}(\prog_2, \nlinput, \mapv) \\
& \text{\ \ \ \ \ in in-order traversal of NLParse(\nlinput)}\\
-1 & \ \ \text{otherwise}\\
\end{array}
\right.
\]
\end{definition}

Features $\featurelcaone, \featurelcatwo$ and $\featureorder$
are used to learn the correspondence between the parse tree
structure and the program structure.  We use these to
maintain the structure of translation close to the structure
of the parse tree.

\begin{definition}[Overlap]
 The overlap feature captures the possibility that 
two programs are constructed from mixtures of two subtrees in the NL Parse tree:
\[
\featureoverlap = \left\{
\begin{array}{rl}
 1 & \mbox{ if } \text{Span}(\prog_1, \mapv_1).\text{end} < \text{Span}(\prog_2, \mapv_2).\text{start}\\
-1 & \mbox{ if } \text{Span}(\prog_1, \mapv_1).\text{start} > \text{Span}(\prog_2, \mapv_2).\text{end}\\
 0 & \text{ otherwise} \\
\end{array}
\right. \\ 
\]
\end{definition}

\begin{definition}[Distance]
Given two programs $\prog_1$ and $\prog_2$ we define the distance feature for programs \ignore{that 
do not overlap} by looking at the distance between the word spans used in the programs:
\[
\featuredistance = \left\{
\begin{array}{ll}
\text{Span}(\prog_2, \mapv_2).\text{start} - \text{Span}(\prog_1, \mapv_1).\text{end} & \mbox{if }\featureoverlap = 1\\
\text{Span}(\prog_1, \mapv_1).\text{start} - \text{Span}(\prog_2, \mapv_2).\text{end} & \mbox{if }\featureoverlap = -1\\
0 & \text{otherwise} \\
\end{array}
\right.
\]
\end{definition}

The features $\featureoverlap$ and $\featuredistance$
capture the proximity information of words and are useful because
related words often occur together in the input sentence.


\begin{example}
Consider possible translations for an input $\nlinput$: \\
\noindent{\small
$\begin{array}{@{}r@{}cl}
\nlinput &:& \mbox{Print all lines that do not contain ``834''} \\
\prog_1 &:& \mbox{PRINT(SelectStr(LINETOK,
  NOT(CONTAINS(} \\ 
  && \qquad \mbox{STRING(834))), ALL()), DOCUMENT())}\\
\prog_2 &:& \mbox{PRINT(SelectStr(STRING(834),
  NOT(CONTAINS(} \\
  & & \qquad \mbox{LINETOK)), ALL()), DOCUMENT())} 
\end{array}
$}\\
In the parse tree NLParse($\nlinput$), ``print'' will have
two arguments, what to print (``lines'') and when to print(``not contain $834$''). We observe the following for the
candidate programs: (a) The word ``lines'' is closer to ``print'', while the
  word ``834'' is farther in NLParse($\nlinput$). The same
  structure is observed for $\prog_1$, but
  not for $\prog_2$. This is captured by LCA Distances. 
(b) The order of the words in NLParse($\nlinput$)
  matches the order in $\prog_1$ than in $\prog_2$. 
  This is captured by the Order feature. 
(c) The phrase ``do not contain 834'' is kept intact in
  $\prog_1$, but is split apart in $\prog_2$. Overlap and Distance
  features will capture this splitting and reordering.

Both the programs use the same set of words, and the same
word to terminal mappings, resulting in the same coverage
score and the same mapping scores. However, the program
$\prog_1$ is correct and our choice of features rank it 
higher.

\end{example}

\begin{table*}[t]
\centering
\scalebox{.93}{\small
\renewcommand{\arraystretch}{1}
\begin{tabular}{|@{\ }c@{}|p{13.0cm}@{\ }|@{\ }p{13mm}@{}|p{12mm}@{}|p{13mm}@{}|@{\ }p{8mm}|} \hline
  & {\bf Program Generated} & {\small\bf Coverage Score} &
  {\small\bf Mapping Score} & 	{\small\bf Structure Score} &
  {\small\bf Final Score}\\ \hline
$\prog_1$ & INSERT(STRING(*), START, IterScope(LINESCOPE, CONTAINS(STRING(P.O. BOX)), ALL))	& 8.33 & 5.73 & 4.45  & 322.17 \\
$\prog_2$ & INSERT(STRING(*), START, IterScope(LINESCOPE, ALWAYS, ALL)) &	5.00 &	8.40 &	4.45 &	248.17 \\
$\prog_3$ & INSERT(STRING(*), START, IterScope(LINESCOPE, STARTSWITH(STRING(P.O. BOX)), ALL)) &	6.67 &	6.35 &	1.09	 & 232.57 \\
$\prog_4$ & INSERT(STRING(P.O. BOX), START, IterScope(LINESCOPE, CONTAINS(STRING(*)), ALL)) &	 8.33 &	5.73 &	1.00 &	272.43 \\
$\prog_5$ & INSERT(STRING(*), START, DOCUMENT)  &	3.33 &	4.74 &	6.84 &	216.33 \\
\hline
\end{tabular}}
\caption{Ranking the set of consistent programs generated by the $\bag$ algorithm.}
\label{tab:example}
\end{table*}

\subsection{Combined Score Example}

To provide some intuition into the complementary strengths
and weaknesses of the various scores, we examine how they
behave on a subset of the programs generated by the $\bag$
algorithm for the following text editing task: \t{Add a
  ``*'' at the beginning of the line in which the string
  ``P.O. BOX'' occurs.}
\autoref{tab:example} shows some of the consistent programs
generated by the $\bag$ algorithm. The first program
($\prog_1$) is the intended translation. Let us look at the performance of each of the component
scores: 

\noindent{\bf Coverage Score}: Both $\prog_1$ and $\prog_4$ use
the maximum number of words from the sentence, and are tied on top
score. $\prog_4$ is  a wrong program as it attempts
to add ``P. O. BOX'' at the beginning of the line containing
``*''.

\noindent{\bf Mapping Score}: The classifier
learnt by our system maps the word ``beginning'' to the
terminal STARTSWITH with a high probability but to the
terminal START with a lower probability. Further, it maps
``occurs'' to the terminal CONTAINS with a still lower
probability. $\prog_2$ does not use the word ``occur'',
otherwise it has same mappings as $\prog_1$.  As a result it
has higher mapping score than $\prog_1$, but suffers on
coverage. $\prog_3$ maps ``beginning'' to STARTSWITH, and
does not use the word ``occurs''. As a result it has a
mapping score lower than $\prog_2$ but higher than
$\prog_1$. If we had used the mapping
score alone, we would not have been able to rank the desired
program $\prog_1$ above the incorrect programs $\prog_3$ and
$\prog_4$.

\noindent{\bf Structure Score}:  Coverage
score and mapping score look only at the mapping of a word 
but not its relation to other mappings and their placement
with respect to the original sentence. Structure score
fixes this by considering structural information (parse
tree, ordering of words and distance among words) from the sentence. 
$\prog_4$ has poor structure score
because it swaps the sentence ordering for strings ``*''
and ``P.O. BOX''. 
$\prog_3$ also suffers as it moves ``beginning'' (mapped to
STARTSWITH) away from ``Add'' (mapped to
INSERT). $\prog_1$ gets a high structure score as it maintains
the parse tree structure of the input text.
Note that, $\prog_2$ and $\prog_5$ have high structure
score as well. This is because structure score does not
take into account the fraction of used words or
word-to-terminal mappings. So, an incomplete translation that uses 
very few words but maps them to correct terminals and
places them correctly, is likely to have a high value.

The desired program, $\prog_1$, is only top ranked by one of
the scores and even in that case, the score is tied with another
incorrect result. However, a combination of the scores
with appropriate weights (\sectref{learning}) \ignore{of 26.91
(Coverage Score), 5.89 (Mapping Score), and 14.43 (Structure
Score),} ranks $\prog_1$  as the clear winner!

\section{Training Phase} \label{sec:learning}
This section describes the learning of classifiers, weights, and the 
word-to-terminal mapping used by the synthesis algorithm described in \sectref{algorithms}. 
The key aspects in this process are (i) deciding which machine learning
algorithm to use, and (ii) generation of (lower level) training data for that machine
learning algorithm from the top level training data provided by the DSL designer. 

\subsection{Mapping Score Classifier ($\classm$)}
\label{sec:learnmapping}
The goal of the $\classm$ classifier is to predict the likelihood of a word 
$\word$ mapping to a terminal $\termexp \in \grammar_T$ 
using the POS tag of the word $\word$. The learning of this classifier is
performed using an off-the-shelf implementation of a Naive Bayesian
Classifier~\cite{bishop}. The training data for this classifier is generated as shown in
\autoref{algo:comp2Train}.
\begin{algorithm}[t]
  \caption{Learning Mapping Score Classifier $\classm$}\label{algo:comp2Train}
  \KwIn{Training Data $\td$}
  \ForEach {training pair $(\nlinput, \prog) \in \td$}{
    $\tilde{\mapv} \gets \mbox{\t{WitnessMaps}}(\prog,\nlinput)$\;
    $\mapv \gets argmax_{\mapv'\in\tilde{\mapv}}(\lst(\prog,\nlinput,\mapv'))$\;
    \ForEach {$(\word,\terminal) \in \mapv$}{
      $\classm$.Train($\word$,POS($\word$,$\nlinput$),$\terminal$)
    }
  }
  \Return $\classm$\;
\end{algorithm}

\newcommand\child{n}
The key idea is to first construct the set $\tilde{\mapv}$ of all witness maps that can
yield program $\prog$ from natural language input $\nlinput$. We then
select the most likely map $\mapv$ out of these witness maps based on 
the partial lexicographic order given by the {\em likeability score} tuples. 
\[\begin{array}{@{}l@{}l@{}l}
\lst(\prog,\nlinput,\mapv) &=& (\UsedWords(\nlinput,\mapv), \\
&& \hspace*{10mm}\djs(\prog,\nlinput,\mapv)) \\
\djs(\prog,\nlinput,\mapv)  &=&  \sum\limits_{\prog' \in  \subprogs(\prog)} \sigma(\prog')\\
\mbox{ where } \sigma((\prog_1,\ldots,\prog_n))  &=&  1
\mbox{ if } \forall \prog_i, \prog_j,  \prog_i \cap \prog_j =
\emptyset,\  0 \mbox{ otherwise }
\end{array}
\]
The likeability tuples serve two purposes: First, via the
\emph{UsedWords}, they guide the system to prefer mappings
that use all parts of the input sentence. Second, via the
\linebreak\emph{\djs}, they guide the system to prefer mappings that
penalize the use of a single part of a sentence to construct
multiple different subprograms.

\subsection{Structure Score Classifiers ($\classs$)}
In this section, we describe how the classifiers used in structure score, $\classs[\conn]$ 
for each connection $\conn$, are learned. 
The goal of each classifier $\classs[\conn]$ is to predict the
likelihood that a combination $\comb$ is an instance of connection
$\conn$ using the $7$ features of $\comb$ from \sectref{ranking}. 
We use an off-the-shelf implementation of a Naive Bayesian
Classifier and generate the training data for it as shown in
\autoref{algo:comp3Train}.

\begin{algorithm}[t]
  \caption{Learning Structure Score Classifiers $\classs$}\label{algo:comp3Train}
  \KwIn{Training Data $\td$}
  \ForEach {training pair $(\nlinput,\prog) \in \td$} {
    $\emph{AllOpts} = \text{SynthNoScore}(\nlinput)$\;
    \ForEach {program $\prog' \in \text{AllOpts}$}{
      \ForEach {combination $\comb$ that occurs in $\prog'$}{
        $\!\!\!$\lIf{$c$ occurs in $\prog$}{class $\gets$ 1}
        $\!\!\!$\lElse{class $\gets$ 0}
        $\!\!\!\conn$ $\gets$ connection used by $\comb$\;
        $\!\!\!\fVector\!\! \gets\!\! {\langle \featureposone, \featurepostwo, \featurelcaone, \featurelcatwo, \featureorder, \featureoverlap, \featuredistance \rangle}$\;
        $\!\!\!\classs$[$\conn$].Train($\fVector$, class);}
    }
  }
  \Return $\classs$
\end{algorithm}

The key idea is to run the synthesis algorithm without the scoring step, \emph{SynthNoScore},  
to construct the set of all programs, \emph{AllOpts}, that can be constructed from the English sentence $\nlinput$. Any 
combination present in a program in $\prog'$ in \emph{AllOpts} but not
present in $\prog$ is used as a negative example, while that
present in $\prog$ is used as a positive example.

\subsection{Dictionary Construction}
\label{sec:dictcons}
We construct the dictionary $\WordToTermMap$ in a semi-automated manner using the names of the terminals 
(functions and arguments) in the DSL. If the name of an operation is a proper English word, such as \t{INSERT},  
we use the \emph{WordNet}~\cite{wordnet} synonym list to gather commonly used words which 
are associated with the action. Cases where the name is not a simple word but instead concatenations of (or 
abbreviations of) several words, such as \t{STARTSWITH}, are handled by splitting the name and resolving 
the synonyms of each sub-component word. 

It is possible that the general purpose synonym sets provided by WordNet contain English words that are not 
useful for the particular domain we are constructing the translator for. However, the mapping score 
learning in \sectref{learnmapping} will simply assign these words low scores. Once the learning algorithm 
for the mappings has finished assigning weights to each word/terminal we discard all mappings below a 
certain threshold. 
Conversely, it is also possible 
that an important domain specific synonym will not be provided by the WordNet sets or that the names in 
the DSL are not well matched with proper English words. Our system automatically detects these cases
as a result of being unable to find witness maps for programs (in the training data)
involving certain DSL terminals. In these cases, it prompts the user to 
identify a word in an input sentence that corresponds to an unmapped terminal in a program. 
These new seed words are then further used to build a more extensive synonym set using WordNet.

\subsection{Learning Combination Weights}
\label{sec:learning-combination}
In the previous section, we defined $3$ component scores for a
translation. A standard mechanism for combining multiple scores 
into a single final score is to use a weighted sum of the component 
scores. In this section we describe a novel method for learning the required 
weights to use to maximize the following function.\\[2mm]
\noindent
\emph{{\bfseries Optimization Function:} Number of benchmarks in the training set, for
which the correct translation is assigned rank 1.}\\[2mm]
In numerical optimization maximization of an optimization function
is a standard problem which can be solved using 
\emph{stochastic gradient descent}~\cite{numericopt}. In order to
use gradient descent to find the weight values that maximize our optimization
function we need to define a continuous and differentiable \emph{loss function},
$F_{\emph{loss}}$. This loss function is used to guide the iterative search for
a set of weights that maximizes the value of the optimization function as
follows:
\begin{center}
 $\vec{\word_{n+1}} = \vec{\word_n} - \gamma \vec{\bigtriangledown}F_{\emph{loss}}(\vec{\word_n})$
\hspace{2 mm} $n=0,1,2,..$
\end{center}
where $\vec{\bigtriangledown}$ denotes the gradient and $\gamma$ is a positive
constant. At each step, $\vec{\word}$ moves in the direction in which the value
of $F_{\emph{loss}}$ decreases and the process is stopped when the change in the
function value in successive steps drops below a specified threshold $\epsilon$.

A common form for loss functions is a sigmoid. We
can convert our ill-behaved optimization function into a loss function that is
closer to what is needed to perform gradient descent by basing the sigmoid on
the ratio score given to the best incorrect result and the score given to the
desired rate via the following construction:
\[
\begin{aligned}
&F_{\emph{loss}}(\vec{\word}) = \sum\limits_{\forall \text{ training } \nlinput}f(\vec{\word}, \nlinput)\\
&f(\vec{\word}, \nlinput) = \frac{1}{1+e^{-c(\lambda-1)}} \text{ where } \lambda=\frac{v_{\emph{wrong}}}{\text{Score}(\prog_{\emph{desired}})} \land c > 0\\
&\quad \prog_{\emph{desired}} = \mbox{\emph{correct translation of }} \nlinput\\
&\quad v_{\emph{wrong}} = \emph{max}(\{\text{Score}(\prog)|\prog\in Bag(\nlinput)\ \wedge\ \prog\neq \prog_{\emph{desired}} \})
\end{aligned}
\]

Although the above transformation results in a loss function
which is mostly
well behaved, it saturates appropriately and is piecewise continuous and
differentiable, there are still points were the function is not continuous. In
particular the presence of the \emph{max} function in the definition of
$v_{\emph{wrong}}$ creates discontinuous points in $F_{\emph{loss}}$. 
However, the following insight enables us to replace the discontinuous
\emph{max} operation with a continuous approximation:
\[
\text{max}(a, b) \approx \text{log}(e^{ca} + e^{cb}) / c \text{ where } c \geq 1 \text{ if } a \ll b \lor b \ll a\\
\]
Thus, we can replace the \emph{max} operator with this function, extended in the
natural way to $k$ arguments, in the computation of $v_{\emph{wrong}}$ to
produce a globally continuous and differentiable loss function. The cases where
there are several incorrect results which are given very similar scores are
minimized by the selection of a large value for $c$, which amplifies small
differences. Additionally, in the worst case where two scores are extremely
close, the impact of the approximation is to drive the gradient descent to
increase the ratio between $v_{\emph{wrong}}$ and
$\text{Score}(\prog_{\emph{desired}})$. Thus, the correctness of the gradient
descent algorithm is not impacted.

In addition to satisfying the basic requirements for performing
gradient descent, 
our loss function, $F_{\emph{loss}}$, saturates for large values of $\lambda$. 
This implies that if an input $\nlinput$, has $\frac{v_{\emph{wrong}}}{\text{Score}(\prog_{\emph{desired}})} \gg 1$ 
it will not dominate the gradient descent causing it to improve the ranking results for 
a single benchmark at the expense of rank quality on a large number of other benchmarks. 
The saturation also implies that the descent will not become stuck trying to find weights 
for an input where there is no assignment to the weights that will improve the 
ranking, i.e., there is an incorrect result program $\prog_i$ where every component 
score has a higher value than the desired program $\prog_d$.

\section{Experimental Evaluation}\label{sec:results}
The (online) synthesis algorithm, consisting of the
\bag\ algorithm and feature extraction (for ranking), was
implemented in C\# and used the Stanford NLP Engine (Version
2.0.2) \cite{stanfordnlpcode} with its default configuration
for POS tagging and extracting other NL features.  The
offline gradient decent was implemented in C\# while the
classifiers used for training the component features were
built using MATLAB.

A major goal of this research is the production of a generic
framework for synthesizing programs in a given DSL from
English sentences. Thus, we selected three different
categories of tasks, question answering (Air Travel
Information System), constraint based model construction
(Automata Theory Tutoring), and command execution on
unstructured data (Repetitive Text Editing).  \ignore{The
  selection of repetitive text editing domain was also
  inspired by recent interest in PBE approaches to text
  manipulation and the challenges of extending example based
  approaches to conditional and repetitive operations.}
These domains, described in detail in~\sectref{motivation},
present a variety of structure in the underlying DSL, the
language idioms that are used, and the complexity of the
English sentences that are seen. For benchmarks, 
{automata descriptions
  are taken verbatim from textbooks and online
  assignments. Text editing descriptions are taken verbatim
  from help forums and user studies. ATIS descriptions are
  part of a standard suite. Tables~\ref{tab:textediting}(b),
  \ref{tab:textediting}(c), \ref{tab:automata} and
  \ref{tab:atis} describe a sample of these benchmarks.} 

\paragraph{Air Travel Information System (ATIS).}
\ignore{The ATIS domain is a well frequently used benchmark
  for natural language understanding research. The
  specialized query DSL we constructed for these benchmarks
  is based around SQL style row/column operations and
  predicates which correspond to important concepts in travel
  queries, arrival/departure locations, times/dates, prices,
  etc.}  We selected $535$ queries at random from the full
ATIS suite (which consists of few thousand queries) and, by
hand, constructed the corresponding program in our DSL to
realize the query. Each task in ATIS domain is a query over
flight related information.

\paragraph{Automata Theory Tutoring.}  \ignore{The automata
  theory domain is motivated by recent work on techniques for
  automating various teaching activities. The specialized
  constraint specification DSL we constructed for these
  benchmarks includes quantification, iterative constructs,
  and constraint predicates. For the NL inputs} We collected
$245$ natural language specifications (accepting conditions)
of finite state automata from books and online courses on
automata theory.

\paragraph{Repetitive Text Editing.}  \ignore{The
  specialized text editing DSL we constructed has multiple
  iteration scopes (lines, words, etc.), compound Boolean
  conditions, a range of commonly used string predicates, and
  commonly used string operations.}  We collected a
description of 21 text editing tasks from Excel books and
help forums. We collected 265 English descriptions for these
21 tasks via a user study, which involved 25 participants
(who were first and second year undergraduate students). The
large number of participants ensured variety in the English
descriptions (e.g., see~\autoref{tab:textediting}(c)). In
order to remove any description bias, each of these tasks was
described not using English but using representative pairs of
input and output examples.  Additionally, we obtained 227
English descriptions for 227 text editing tasks (one for each
task) from an independent corpus~\cite{acl11}.

\subsection{Precision, Recall, and Computational Cost}
In this study we used standard \emph{10-fold
  cross-validation} to evaluate the precision and recall of
the translators on each of the domains.  Thus, we select
$90\%$ of the data at random to use for learning the
classifiers/weights and then evaluate the system on the
remaining $10\%$ of data which was held back (and not seen
during training). In the ranking we handle ties in the scores
assigned to an element using a \emph{1334 ranking}
scheme~\cite{wikipedia-ranking}. In 1334 ranking, in the case
of tied scores, each element in the tied group is assigned a
rank corresponding to the lowest position in the ordered
result list (as opposed to the highest).  This ensures that
the reported results represent the \emph{worst case} number
of items that may appear in a ranked list before the desired
program is found.

\begin{figure}
    \includegraphics[scale=0.4]{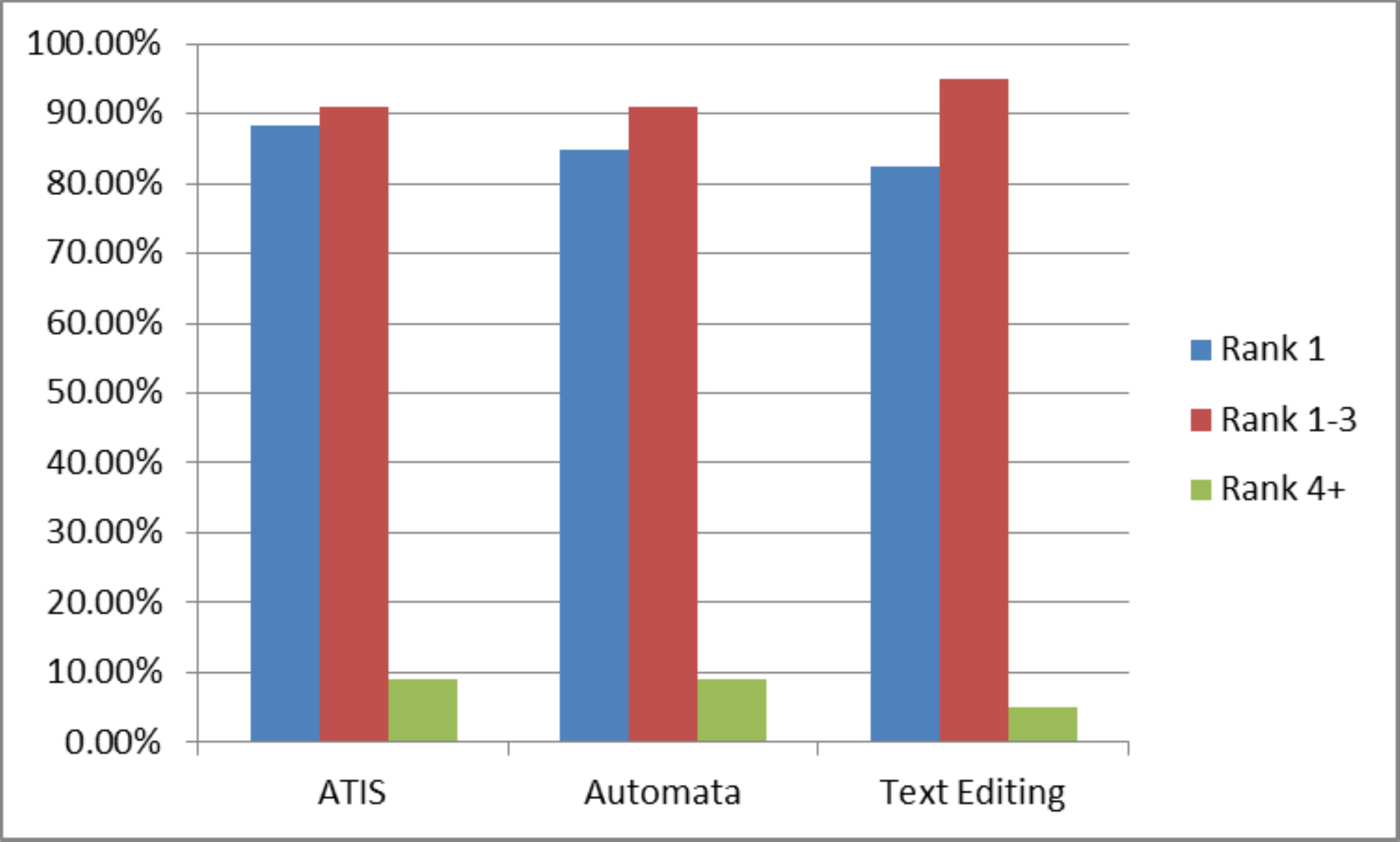}
    \caption{\label{fig:rankresults}Ranking precision of
      algorithm on all domains.}
\end{figure}

\paragraph{Precision.}
\autoref{fig:rankresults} shows the percentage of inputs for which the desired
program in the DSL is the top ranked result, the
percentage of inputs where the desired result is in the first three results, and
the percentage where the desired result may be more than three entries down in
the result list. As shown in the figure, for every domain, on over $80$\% of the
inputs the desired program is unambiguously identified as the top ranked
result. Further, for the ATIS domain the desired result is the top ranked result
for $88.4$\% of the natural language inputs. Given the size of our sample from
the full ATIS suite we can infer that the desired program will be the top
ranked result for $88.4\pm4.2$\% of the natural language inputs at a $95$\%
confidence interval. These results show that our novel \emph{program synthesis} based 
translation approach is competitive with \ignore{or better
  than} the state-of-the-art \emph{natural language processing} 
systems: 85\% in \citet*{Zettlemoyer}, 84\% in \citet*{Hoifung}, and 83\%
in \citet*{Kwiatkowski}.

\paragraph{Recall.}
In addition to consistently producing the desired program as the top ranked
result for most inputs the ranking algorithm places the desired program in the
top $3$ results an additional $5$\%-$12$\% of the time. Thus, across all three
of the domains, for over $90$\% of the natural language inputs the desired
program is one of the three top ranked results. This leaves less than $10$\% of
the inputs for any of the domains, and only $5$\% in the case of the text
editing domain, where the synthesizer was unable to produce and place the desired
program in the top three spots.

\begin{figure}
  \begin{center}
    \includegraphics[scale=0.66]{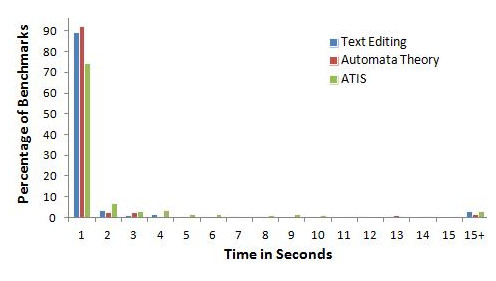}
    \caption{\label{fig:times}Timing performance of algorithm on all domains.}
  \end{center}
\end{figure}

\paragraph{Computational Cost.}
\autoref{fig:times} shows the distribution of the time
required to run the synthesis algorithm and perform the
ranking. On average translation takes $0.68$ seconds for Text
Editing, $1.72$ seconds for Automata and $1.38$ seconds for
the ATIS inputs. Further, the distribution of times is
heavily skewed with more than $85$\% of the inputs taking
under $1$ second and very few taking more than $3$
seconds. The outliers tend to be inputs in which the user has
specified an action in an exceptionally redundant manner.

\subsection{Individual Component Evaluation}
In \sectref{algorithms} we defined various components for ranking and provided 
intuition into their usefulness. To validate that these component scores are, in fact, 
important to achieving good results we evaluated our choices by using various subsets of 
the component scores, learning the best weights for the
subset, and re-ranking the programs. 

\paragraph{Performance of Individual Scores.}
The results of using each component in isolation are
presented in \autoref{res:weakness of components}.  This
table shows that when identifying the top-ranked program the
best performance for using only \covs\ is $17.9$\%, using
\maps\ is $31.0$\% and for \strs\ is $51.4\%$. This is far
worse than the result obtained by using the combined ranking
which placed the desired program as the top result for
$84.9\%$ of the inputs. Thus, we conclude that the components
are not individually sufficient.

\begin{table}
  \begin{center}\small
    \begin{tabular}{|@{}l@{}|@{}c@{}|@{}c@{}|@{}c@{}|}
      \hline
      & \multicolumn{3}{c|}{\bf 1334 Top Rank} \\ \cline{2-4}
      \headstart{Domain} &  \head{\covs} &
      \head{\maps} & \head{\strs} \\ \hline
      \domATIS & 5.6\% &  1.9\% &  20.2\%  \\ \hline
      \domAut & 17.9\% &  31.0\% & 51.4\%  \\ \hline
      \domTE & 8.1\% & 8.9\% &  34.4\% \\ \hline
    \end{tabular}
  \end{center}
  \caption{Performance of individual component scores in ranking the desired program as top result.}
  \label{res:weakness of components}
\end{table}

\paragraph{Score Independence.}
Although these results show that independently none of the components are
sufficient for the program ranking it may be the case that one of the components
is, effectively, a combination of the other two. \autoref{res:significance of
components} shows the results of ranking the programs when dropping one of the
components. Dropping \strs\ results in the largest decrease, as high as
$81.86$\% in the worst case, and even the best case has a decrease of $47.75$\%.
Dropping \covs\ also results in substantial degradation, although not as high as
for \strs. The impact of dropping \maps\ is much smaller, between $2.04$\% and
$4.67$\%. However, the consistent positive contribution of \maps\ shows that it 
still provides useful information for the ranking. Thus, all of the components 
provide distinct and useful information.

\begin{table}
\begin{center}\small
  \begin{tabular}{|@{}l@{}|r|r|r|}
  \hline
  & \multicolumn{3}{c|}{\% change on dropping} \\ \cline{2-4}
   \headstart{Domain} &  \head{\covs} &
\head{\maps} & \head{\strs} \\ \hline
  \domATIS & -30.6\% & -4.7\% & -81.9\% \\ \hline
  \domAut & -22.4\% & -2.0\% & -47.7\% \\ \hline
  \domTE & -19.7\% & -4.7\% & -65.8\% \\ \hline
  \end{tabular}
\caption{Impact of dropping individual component scores on top rank percentage.}
\label{res:significance of components}
\end{center}
\end{table}

\paragraph{Dictionary Construction.}
In practice the semi-automated approach makes dictionary construction a task that, 
while usually requiring manual assistance, does not require expertise in natural language 
processing or program synthesis. On average the dictionaries for each domain 
contained 144 English words averaging 4.51 words/terminal and 1.48 terminals/word. 
The user was prompted to provide 20.7\% mappings on average
across the three DSLs. Although beyond the scope of this work, as it requires a larger corpus of training 
data, the amount of user intervention can be further reduced by using statistical alignment
to automatically extract the domain specific synonyms from the training data.

\paragraph{Score Combination Weights.}
We used gradient descent to learn how much to weight each score in the computation of 
the final rank of a program. To evaluate the quality of the weights identified via the 
gradient descent we compared them with a naive selection of equal weights for all 
the component scores and with the results of boosting. Boosting~\cite{RankBoost} is a 
frequently technique which combines a set of weaker rankings, such as the individual 
component scores, to produce a single strong ranking. \autoref{res:effect of gradient descent} 
shows the results of the rankings obtained with the three approaches.

\begin{table}
\begin{center}\small
 \begin{tabular}{|@{}l@{}|@{}c@{}|c@{\%}|@{}c@{\%}|c@{\%}|@{}c@{\%}|c@{\%}|@{}c@{\%}|}
   \hline
   \headstart{Domain} & \head{Total} &
   \multicolumn{2}{@{}c@{}|}{\bf Equal Wt.} & \multicolumn{2}{@{}c@{}|}{\bf Rank Boost} &
   \multicolumn{2}{@{}c@{}|}{\bf Gradient}\\ \cline{3-8}
   & \head{Count} & \head{Top} & \head{Top-3} & \head{Top} & \head{Top-3} & \head{Top} & \head{Top-3} \\
   \hline
   \domATIS & 535 & 73.2 & 90.8 & 79.8 & 89.9 & 88.4 & 93.3 \\ \hline
   \domAut  & 245 & 74.2 & 91.4 & 73.1 & 93.5 & 84.9 & 91.2 \\ \hline
   \domTE   & 492 & 74.0 & 91.1 & 74.4 & 91.8 & 82.3 & 94.9 \\ \hline
 \end{tabular}
 
 \caption{Comparison of ranking using equal weights, gradient
   descent, and RankBoost. Column ``Top''(``Top-3'') shows the percentage of
   benchmarks where the correct translation is ranked 1
   (ranked in top 3).}
   \label{res:effect of gradient descent}
\end{center}
\end{table}

The results show that using gradient descent has improved the number of top ranked 
benchmarks significantly over the naive weight selection (as large as $15$\%). However, the improvement in the top 3 
ranked benchmarks is much smaller. Similarly, the gradient descent approach produces 
substantially better results than RankBoost with an average difference of $9$\% in the top 
ranked benchmarks. Thus, we can conclude that the use of gradient descent for learning 
the combination weights is an important factor in the overall quality of the results.

{Our choice of the ranking functions is critical to the quality of results. As shown in \autoref{res:significance of components}, dropping any of the component functions results 
in a substantial loss of precision. 
Also, using a 
simpler method, such as equal weights or boosting~\cite{RankBoost}, 
to compute the combination weights results in a loss of 9-15\% 
in precision when compared to the use of gradient descent (\autoref{res:effect of gradient descent}).}

{In our system, most failures (i.e. the correct solution failing to rank in the top-three solutions) arise because some key information is left implicit
in the English description, e.g. ``I want to fly to Chicago
on August 15''. In this case, the departing city should default to
``CURRENT\_CITY'' \ignore{ instead of ``ANY''}, and the time should default to ``ANY''
\ignore{instead of ``CURRENT\_TIME''}.\ignore{, and the date (August 15) should be taken from 
the users input.} Such issues might be fixed either by
having orders of magnitude larger training data or by building some
specialized support for handling implicit contextual information in
various domains.}

\begin{figure}
\includegraphics[scale=0.5]{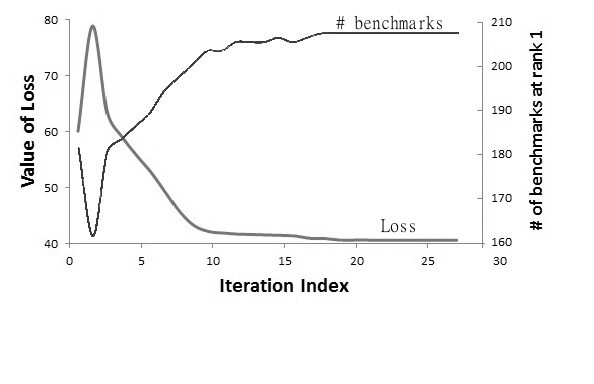}
\vspace{-12mm}
\caption{Behavior of Loss Function and Top Ranks.}
\label{fig:loss behavior}
\vspace{-3mm}
\end{figure}

As part of learning the weights for the component scores we used a 
shifted variant of the logistic function as our loss function (\sectref{learning}). 
\autoref{fig:loss behavior} shows how the value of loss changes with
iteration index and the corresponding number of top ranked benchmarks. 
It can be seen that as the loss value decreases, the number of top ranked benchmarks increases
and vice-a-versa. Thus, as these values are negatively correlated as needed for optimal 
performance of the gradient descent algorithm, and even though our loss function contains 
the log-exponential approximation of the \emph{max} operation it is well behaved for the 
gradient descent algorithm.

Since the weights learned in~\sectref{ranking} are general
purpose, we expect that weights learned from one domain are
applicable to other domains, eliminating the
the time and effort required to re-learn these values on each
new domain.  The results in table~\ref{res: generality}
show that the weight vectors that are learned for one domain
perform well when used to rank the results for a new domain.
The average decrease in the number of top ranked programs is
only $1.9$\% (with a maximum decrease of $5.0$\%). For the
number of top 3 ranked programs the change is insignificant
with a maximum decrease of less than $0.5$\% and thus we do
not include them here. This result demonstrates that the
learning of the component weights is highly domain
independent and generalizes well, allowing it to be reused
(or used as a starting point) for new domains.  \cmt{This
  domain-independence in learning can be attributed to the
  'careful' design of grammars and the choice of
  features. The grammars are designed in such a way that the
  terminals are almost in one-to-one mapping with English
  language words. The combination of terminals in top-ranked
  pogroms closely follow the English sentence because the
  intuition behind the three main scores for ranking comes
  from the structure and rules of English language.  So,
  irrespective of the domain, the learned weights fit nicely
  across the domains.}

\begin{table}
\small
\vspace{-3mm}
\begin{center}
 \begin{tabular}{|@{}l@{}|r|r|r|}
  \hline
   & \multicolumn{3}{c|}{\bfseries \% change on using weights learnt for}\\
   \headstart{Domain} & \head{\domATIS} &
\head{\domAut} & \head{\domTE}\\
  \hline
\domATIS & 0.0\% & -1.5\% & -5.0\% \\ \hline
 \domAut & -2.0\% & 0.0\% & -1.2\% \\ \hline
\domTE & -0.6\% & -0.8\% & 0.0\%  \\ \hline
 \end{tabular}
\caption{Generalization of learning across domains.}
\label{res: generality}
\vspace{-3mm}
\end{center}
\end{table}


\section{Related Work}

\balance

\paragraph{PBE/PBD Techniques for Data Manipulation} 
Programming by demonstration (PBD) systems, which use a
 trace of a task
performed by a user, and programming by example (PBE) systems, which learn from
a set of input-output examples, have been used to enable end-user programming
for a variety of domains. For PBD these domains include text
manipulation~\cite{smartedit} and table transformations~\cite{wrangler} among
others~\cite{cypher-pbd}. Recent work on PBE by Gulwani et. al. has included
domains for manipulating strings~\cite{popl11,vldb12}, numbers~\cite{cav12}, and
tables~\cite{harris:pldi11}. As mentioned earlier, both PBD and PBE based techniques struggle when the
desired transformations involve conditional operations.\ignore{ One issue is potential
exponential growth in the number of examples needed when learning conditions,
where examples exercising both the true and false paths are needed. A second
issue is the difficulty of generalizing from user examples, which often under
specify the desired condition, to the precise condition which the user intended.}
In contrast the natural language based approach in this work performs well for
both simple and conditional operations.

\paragraph{Keyword Programming}
Keyword programming refers to the process of translating a set or sequence of
keywords into function calls over some API. This API may consist either of
operations in an existing programming language~\cite{ase07,pldi12,jungloid2005}
or a DSL constructed for a specific class of tasks~\cite{koala,chickenfoot}.
Keyword programming techniques use various program synthesis approaches to build
expression trees from the elements of the underlying API, similar to the \bag\
algorithm in~\sectref{algorithms}, and then use simple heuristics, such as words
used and keyword-to-terminal weights, to rank the resulting expression trees. These 
systems have low precision and, as a result, will frequently suggest incorrect programs.

\paragraph{Semantic Parsing}
Semantic parsing~\cite{semparsingreview} is a
sophisticated means of constructing a program from natural language using a
specialized language parser. Several approaches including, syntax directed~\cite{semparsingsyntax}, NLP
parse trees~\cite{semparsingtree}, SVM driven~\cite{semparsingsvm}, combinatory
categorical grammars~\cite{zettlemoyer05learning,Zettlemoyer,Kwiatkowski}, and dependency-based
semantics~\cite{liang11dcs,Hoifung} have been proposed.
These systems have high precision, usually suggesting the correct program, 
but low recall and often do not return any suggestions at all.
In contrast the technique in this paper achieves similar levels of precision 
but does not suffer from low recall.

\paragraph{Natural Language Based Programming}
A number of natural language programming systems have been built around
grammars, NLC~\cite{nlc}, or templates, NaturalJava~\cite{njava}, which impose
various constraints on input expressions. Such systems are sensitive to grammatical errors or extraneous words.
There has been extensive research on developing natural language interfaces to
databases (NLIDB)~\cite{nlidb:95,nlidb11}. While early systems were based on
pattern matching the user's input to one of the known patterns, 
PRECISE~\cite{precision,precision:iui03} translates {\em semantically tractable}
NL questions into corresponding SQL queries. However, these systems
depend heavily on the underlying data having a known schema which makes them
impractical when the underlying data structure is unknown (or non-existent) as
in the text-editing domain used in this work.

SmartSynth~\cite{mobisys:2013} is a system for synthesizing smartphone scripts
from NL. The synthesis technique in SmartSynth is highly specialized to the
underlying smartphone domain and uses a simple the ranking strategy for the
programs that it produces. Similarly, the NLyze~\cite{nlyze:2014} system
synthesizes spreadsheet formulas from NL. Again, NLyze is designed for a
specific domain (spreadsheet formula) and uses a relatively simple ranking
system consisting of only the equivalent of the coverage, mapping, and overlap
features presented in our paper. In contrast, the system presented in this paper
is agnostic to the specifics of the target DSL, the ranking features are
independent of the underlying DSL, and we automatically learn appropriate
weights for the features. In addition, as the experimental results in
\autoref{res:significance of components} demonstrate, the use of a simpler
ranking system, as in SmartSynth or NLyze, results in substantial reductions in
recall/precision. Thus, the approach in this paper can be seen as an improvement
and generalization of these previous systems.

\section{Conclusion}

{Today billions of end-users have access to computational
  devices, yet lack the programming knowledge to effectively
  interact with these devices. Most of these users are
  wanting to write small programs or specifications that can
  be described succinctly in some appropriate domain-specific
  language that provides the right level of
  abstractions. These users are stuck because of the need to
  provide step-by-step, detailed, and syntactically correct
  instructions to the computer. Program synthesis has the
  potential to revolutionize this landscape, when targeted
  for the right set of problems and using the right
  interaction model.
  
  Programming-by-example has been shown to be a
  very effective tool---a recent instance being release of
  the Flash Fill feature~\cite{popl11} as part of Microsoft
  Excel 2013 among rave reviews~\cite{flashfill}. We observe
  that there are several domains for which examples is not a
  natural form of specification (or where too many examples
  would be required), but those tasks can be easily expressed
  in a natural language.
}

 We presented a novel technology for synthesizing programs from
 natural language descriptions (based on generating and
 ranking programs from a set of terminals that correspond to
 the words in the natural language description). More
 significantly, we showed how this framework 
 allows creating synthesizers for different DSLs by simply
 providing {\em examples} of translations. 

We believe that technique will work with off-the-shelf DSLs
without major modifications provided the DSL is functional
without binding constructs such as temporary variables or
quantifiers and has a level of abstraction with a direct
correspondence to the abstraction being used in the natural
language. For example, our approach will not work well for
translating descriptions for automata construction problems
into a target DSL of regular expressions because there is no
direct correspondence. This lack of correspondence between
the source language and the target DSL requires the
translator to use non-trivial logical reasoning during the
conversion and greatly reduces the effectiveness of our
system.

As with any program synthesis technique which fundamentally
involve search over exponential spaces, the cost of our
technique is also worst case exponential in the size of the
DSL. However, the key issue is doing this efficiently for
practical cases.  Our synthesis
works efficiently (usually under 1 second) for a range of
useful DSLs. The size of the dictionary has minimal impact on
the runtime as the translation only depends on the subset of
the dictionary corresponding to the words in the input
sentence.

In future, we aim to further generalize the framework to
allow synthesis of synthesizers for a wider variety of
domains.

\bibliographystyle{abbrvnat}
\bibliography{paper}

\end{document}